%% file: main.tex
\documentclass[letterpaper,oneside,11pt]{article}

\usepackage[utf8]{inputenc}
\usepackage{amsthm}
\usepackage{amssymb}
\usepackage{amsmath}
\usepackage{geometry}
\usepackage{commath}
\usepackage{graphicx}
\usepackage{subcaption}
\usepackage{bm}
\usepackage{float}
\usepackage{caption}
\usepackage{setspace}
\usepackage{multicol}
\usepackage{color}
\usepackage{subfiles}
\usepackage{textcomp}
\usepackage{enumerate}
\usepackage{hyperref}
\usepackage{commath}
\usepackage{mathrsfs}
\usepackage{comment}
\usepackage{cite}
\usepackage{listings}
\usepackage{booktabs}
\usepackage[english]{babel}
\usepackage{multirow}
\usepackage{algorithm}
\usepackage[noend]{algpseudocode}
\usepackage{nicematrix, tikz-cd, tikz}
\usetikzlibrary{matrix}
\usetikzlibrary{shapes,decorations,decorations.pathreplacing,calligraphy,arrows,calc,arrows.meta,fit,positioning}
\tikzset{
    -Latex,auto,node distance =1 cm and 1 cm,semithick,
    state/.style ={ellipse, draw, minimum width = 0.7 cm},
    point/.style = {circle, draw, inner sep=0.04cm,fill,node contents={}},
    bidirected/.style={Latex-Latex,dashed},
    el/.style = {inner sep=2pt, align=left, sloped}
}  

\usepackage{setspace}
\usepackage[compact]{titlesec}
\titlespacing{\section}{0pt}{*0.8}{*0.8}
\titlespacing{\subsection}{0pt}{*0.8}{*0.8}
\titlespacing{\subsubsection}{0pt}{*0.8}{*0.8}

\newcommand{\bA}{ {\boldsymbol A} }

\newcommand{\bD}{ {\boldsymbol D} }

\newcommand{\bI}{ {\boldsymbol I} }

\newcommand{\bK}{ {\boldsymbol K} }

\newcommand{\bu}{ {\boldsymbol u} }
\newcommand{\bU}{ {\boldsymbol U} }

\newcommand{\bw}{ {\boldsymbol w} }

\newcommand{\bx}{ {\boldsymbol x} }
\newcommand{\bX}{ {\boldsymbol X} }
\newcommand{\by}{ {\boldsymbol y} }
\newcommand{\bY}{ {\boldsymbol Y} }
\newcommand{\bz}{ {\boldsymbol z} }
\newcommand{\bZ}{ {\boldsymbol Z} }

\newcommand{\bbeta}   { {\boldsymbol \beta} }
\newcommand{\bgamma}  { {\boldsymbol \gamma} }

\newcommand{\bepsilon}{ {\boldsymbol \epsilon} }

\newcommand{\bkappa}  { {\boldsymbol \kappa} }
\newcommand{\blambda} { {\boldsymbol \lambda} }

\newcommand{\bmu}     { {\boldsymbol \mu} }

\newcommand{\bSigma}  { {\boldsymbol \Sigma} }

\newcommand{\bPhi}    { {\boldsymbol \Phi} }

\newcommand{\bzero}  { {\boldsymbol 0} }

\def\T{{ \mathrm{\scriptscriptstyle T} }}

\title{Robust Distributed Learning of Functional Data From Simulators through Data Sketching}
\author{
    R. Jacob Andros \\
    Department of Statistics, Texas A\&M University\\
    Rajarshi Guhaniyogi\\ 
    Department of Statistics, Texas A\&M University\\
    Devin Francom \\
    Los Alamos National Laboratories\\
    Donatella Paqualini\\
    Los Alamos National Laboratories
    }
\begin{document}

\maketitle

\begin{abstract}
\noindent In environmental studies, realistic simulations are essential for understanding complex systems. Statistical emulation with Gaussian processes (GPs) in functional data models have become a standard tool for this purpose. Traditional centralized processing of such models requires substantial computational and storage resources, leading to emerging distributed Bayesian learning algorithms that partition data into shards for distributed computations. However, concerns about the sensitivity of distributed inference to shard selection arise. Instead of using data shards, our approach employs multiple random matrices to create random linear projections, or sketches, of the dataset. Posterior inference on functional data models is conducted using random data sketches on various machines in parallel. These individual inferences are combined across machines at a central server. The aggregation of inference across random matrices makes our approach resilient to the selection of data sketches, resulting in \emph{robust distributed Bayesian learning}. An important advantage is its ability to maintain the privacy of sampling units, as random sketches prevent the recovery of raw data. We highlight the significance of our approach through simulation examples and showcase the performance of our approach as an emulator using surrogates of the Sea, Lake, and Overland Surges from Hurricanes (SLOSH) simulator—an important simulator for government agencies.
\end{abstract}

\input{body.tex}

\bibliographystyle{plain}
\bibliography{references}
\end{document}

%% file: body.tex
\section{Introduction}\label{sec:intro}

In environmental applications, scientific analysis often relies on high-resolution spatiotemporal physics-based simulations. For instance, various physics-based simulators for climate and weather systems generate detailed spatial simulations based on input attributes \cite{petersen2019evaluation,borge2008comprehensive}. This manuscript specifically centers on the Sea, Lake, and Overland Surges from Hurricanes (SLOSH) simulator \cite{slosh1992}, developed by the National Weather Service for operational hurricane monitoring and response. Configured with a specific area of interest, SLOSH takes a hurricane's track and a vector of hurricane-related attributes as input, and produces a spatio-temporal grid of storm surge, or a measure of flood depth above normal levels (e.g., water depth above ground level for land spatial locations). In our study, we focus on determining the maximum flood depth at each spatial location for a given hurricane, with emphasis on flooding of electrical substations. Our study area encompasses a part of the Delaware Bay, which separates the southern end of New Jersey from the northern side of Delaware (see Figure~\ref{fig:locations}).

To characterize the spatial variation of maximum flood depth and its relationships with input attributes, as well as to address input uncertainty, one approach is to treat the simulator as a black box and explore it by running it with various settings \cite{sacks1989design}. However, this approach involves conducting numerous runs of the simulator with various settings, and each of these runs comes with substantial computational costs. As a result, the process of simulating multiple runs is significantly expensive and often infeasible in terms of computational resources. A solution to these challenges often involves creating a statistical surrogate model, or emulator, for the simulator. Once trained, these emulators provide rapid predictions for new input settings, facilitating studies of model response and parameter uncertainty.  \cite{hutchings2023storms} compare four emulation methods for SLOSH, though a significant numbers of model runs were necessary for satisfactory performance of the framework. In contrast, the focus of this work is to build a SLOSH emulator with \emph{very few model runs} while taking special care to treat the spatial association of the water depth level and allowing for part of data stored in different centers or servers locally.

Considering flood depth as functional data across spatial locations, Gaussian processes (GPs) can offer a suitable framework for accounting for the correlation between outputs at different locations. GPs have gained prominence as effective predictive tools in the domain of computer experiments, as they can significantly reduce the computational burden associated with running simulations while maintaining flexibility and allowing for comprehensive uncertainty analysis encompassing both parameter and code uncertainties \cite{kennedy2001calibrate,salter2016comparison}. In the specific context of the SLOSH simulator, corresponding to an input vector $\bz_s\in\mathbb{R}^p$ in the $s$th simulation ($s=1,...,S$), we observe noisy outputs $y_{s}(\bu_1),...,y_{s}(\bu_n)\in\mathbb{R}$ from the simulator at $d$-dimensional index points $\bu_1,...,\bu_n\in\mathcal{U}\subseteq\mathcal{R}^d$, respectively. These input attributes remains same corresponding to the output at all index points, and is referred to as \emph{global attributes} throughout the article. We observe an additional $q$-dimensional covariates $\bx(\bu_1),...,\bx(\bu_n)$ at each index point, referred to as \emph{local attributes}. With global and local attributes, one can build a GP emulator described by the following model,
\begin{align}\label{index_model}
y_s(\bu_i)=\bz_s^T\bgamma+\bx(\bu_i)^T\bbeta+w_s(\bu_i)+\epsilon_s(\bu_i),\:\:i=1,...,n,\:\:s=1,...,S,
\end{align}
where $\bgamma$ and $\bbeta$ are $p$ and $q$ dimensional coefficients corresponding to the global and local attributes, respectively. The term $\epsilon_s(\bu_i)$ denotes idiosyncratic errors, assumed to be independently and identically distributed with a Normal distribution with mean $0$ and variance $\tau^2$, for simplicity. The unknown function $w_s(\cdot)$, accounting for variability across the domain $\mathcal{U}$ is considered a realization from a GP with a mean of $0$ and a covariance kernel $\kappa(\cdot,\cdot)$, independently over the simulations $s=1,...,S$.

Performing Bayesian inference for the model (\ref{index_model}) with a Gaussian process prior on $w_s(\bu)$ becomes computationally impractical for a large sample size due to the $O(n^3)$ computational cost and the $O(n^2)$ storage cost associated with estimating $w_s(\bu)$. The domain of modeling high-dimensional dependent functional data has seen substantial growth in the last decade, which has been largely adapted and built upon scalable spatial models. While the extensive literature in this area cannot be fully covered here, comprehensive reviews can be found in \cite{heaton2019case}.

This literature has largely operated within a centralized data processing framework, where all data is stored and processed at a central location. However, an alternative approach involves distributed Bayesian learning for functional data \cite{guhaniyogi2018meta,guhaniyogi2023distributed,guhaniyogi2019multivariate,guhaniyogi2022distributed}. This approach extends the scalability of existing well-established algorithms to estimate (\ref{index_model}), without necessitating the development of new algorithms or software. In essence, this methodology partitions the data into a multitude of shards, wherein a suitable functional data model such as (\ref{index_model}) is fitted to each shard to draw parallel posterior inferences. These individual inferences are then combined at a central server, reducing computation and storage requirements on the server itself. Distributed inference eliminates the need for extensive raw data exchange and communication between the central server and processors during statistical analysis. Consequently, this leads to lower latency and reduced communication traffic. Furthermore, as each processor analyzes a smaller data shard, the computation and storage becomes more efficient, contributing to overall faster inference. Significantly, numerous state-of-the-art methods in the analysis of large functional data come from spatial statistics research, though they crucially use smaller dimensionality of $\mathcal{U}$ in their design, as exemplified by methods that utilize nearest neighbors \cite{datta2016hierarchical}. These methods may not be entirely suitable when $\mathcal{U}$ is higher-dimensional since all points in higher dimensions are approximately equidistant. In contrast, distributed approaches offer scalability without the need to leverage the reduced dimensionality of $\mathcal{U}$, thus making them suitable for the seamless application to functional data models with higher dimensional $\mathcal{U}$. Nevertheless, it is important to note that inference in distributed Bayesian learning can become sensitive to the choice of data subsets, as highlighted by simulation studies in \cite{guhaniyogi2023distributed}.

We will address this issue by introducing a distributed Bayesian inferential framework for functional data modeling that employs Bayesian data sketching techniques \cite{vempala2005random, halko2011finding, mahoney2011randomized, woodruff2014sketching, guhaniyogi2015bayesian, guhaniyogi2016compressed}. The concept of data sketching involves compressing the complete dataset before applying a model, facilitating more efficient computation and storage. While data sketching methods have gained popularity in high-dimensional and penalized regression contexts with large datasets \cite{zhang2013recovering, chen2015fast, dobriban2018new, drineas2011faster, ahfock2017statistical, huang2018near}, their application to resolving computational challenges in Bayesian high-dimensional and functional data regression remains limited, with a few notable exceptions \cite{guhaniyogi2021sketching}. 

%Our approach diverges from the creation of data shards and instead relies on multiple data sketches as the foundation of a distributed Bayesian learning framework. 
The overarching outline of this framework is as follows. First, we generate $H$ random matrices, each with dimensions $m \times n$, where $m$ is significantly smaller than $n$, following the literature on data sketching, and construct $H$ random linear sketches of both the response vector and the predictor matrix. Secondly, we undertake parallel posterior computations on these $k$ data sketches using the functional data model (\ref{index_model}). In this process, the likelihood in the model is raised to the power of $n/m$ for each sketch. This manipulation ensures that the variance of each sketch's posterior distribution is of the same order of $n$ as that of the full data posterior distribution. This set of pseudo posterior distributions, constructed for each data sketch, is referred to as the ``sketched posterior." Thirdly, we calculate the Wasserstein mean of these $k$ sketched posterior distributions, yielding a single probability distribution termed the ``collaborative sketched posterior" distribution. This collaborative sketched posterior serves as a computationally tractable approximation to the full posterior distribution.

The proposed approach addresses a number of important issues on distributed Bayesian learning simultaneously. First, unlike current distributed Bayesian learning methods that perform inference conditional on a fixed choice of data shards, our proposed approach takes a distinct route. Instead of being confined to a single set of data shards, we compute the Wasserstein mean of sketched posteriors obtained using different random matrices. This innovative technique mitigates the sensitivity of inference stemming from the selection of specific random matrix in the construction of data sketches. As a result, we introduce a novel category of distributed learning paradigm, referred to as the ``robust distributed learning," that is resilient to the influences of data shard choices. Second, a crucial aspect of the proposed methodology is its ability to uphold the privacy of sampling units throughout the inference. This is achieved by revealing only lower-dimensional sketches to the analysts which are designed in such a way that the mutual information between them and the full data converges to zero as sample size becomes large, rendering the recovery of the full data from data sketches practically impossible \cite{zhou2008compressed}. Importantly, the framework allows constructing a sketch for the full data from sketches of subsets of data stored privately in different research centers.

%We leverage recent advancements in the fields of random matrix theory and distributed Bayesian learning to analyze the accuracy of approximating the true function using the sketched approximate posterior. Our theoretical analysis reveals that when the parameter $k$ is chosen to increase in a certain proper order with respect to the sample size $n$ as $n$ approaches infinity, \textbf{conclusion of the theoretical result}. While recent theoretical progress has been made in both distributed Bayesian learning with nonparametric models \cite{guhaniyogi2020distributed,guhaniyogi2017divide,szabo2019asymptotic}, as well as in the Bayesian data sketching literature for ordinary high-dimensional linear regression \cite{guhaniyogi2022sketching}, our contribution lies in the establishment of theoretical results for data sketching in the context of distributed Bayesian learning. This novel exploration, to the best of our knowledge, fills a gap in the current research landscape.

The rest of this article progresses as follows. Section~\ref{slosh} describes the data simulated from SLOSH simulator. Section \ref{three_step} describes the robust distributed learning approach in detail. Implementation of the approach on our own simulated data, followed by the SLOSH simulator analysis, are presented in Sections~\ref{simulations} and \ref{data_analysis}, respectively. Finally, Section~\ref{conclude} concludes the paper with an eye towards future work. %Proofs of the theoretical results, as well as additional empirical analysis are described in the supplementary material.
% Section~\ref{theory} offers the theoretical development for the proposed approach. Empirical investigation of the model with a variety of simulation studies.

\section{SLOSH Emulator Data}\label{slosh}
Storm surge models simulate floodwater depth resulting from hurricanes, and are used for emergency response, planning, and research.
The Sea, Lake, and Overland Surges from Hurricanes (SLOSH) simulator (Jelesnianski et al., 1992) is one such model developed by the National Weather Service. Various other storm surge models exist, with more complex models requiring extreme computation time to run. While SLOSH is not prohibitively expensive, our purpose in this paper is to develop emulator for expensive models often encountered in national laboratories that can only be run at a handful of times. Therefore, our dataset comprises an ensemble of only $10$ simulations executed using the SLOSH simulator, representing 10 distinct simulated storms. Each storm within this ensemble is characterized by a unique combination of five input parameters or global attributes. Four of these attributes describe hurricane characteristics when the hurricane makes landfall, including the heading of the eye, the velocity of the eye, the latitude of the eye, and the minimum air pressure experienced.  The fifth attribute is the projected sea level rise for the year 2100, which of course is uncertain. General ranges for these parameters are given in Table \ref{tab:ranges}, from which 10 simulations were generated. % not actually a LHS anymore - perhaps instead of the first 10 we could try to choose a space-filling 5 for training and 5 random for test (more for test would be better).

\begin{table}[]
    \centering
    \begin{tabular}{lccc}
    \toprule[1.5pt]
        Predictor & Lower & Upper & Units \\
        \midrule
       Heading & 204.0349 & 384.0244 & degrees \\
       Velocity & 0 & 40 & knots\\
       Latitude & 38.32527 & 39.26811 & degrees \\
       Pressure & 930 & 980 & millibars \\
       Sea level rise (2100) & -20 & 350 & cm\\
       \bottomrule[1.5pt]
    \end{tabular}
    \caption{Parameters varied in SLOSH simulations.}
    \label{tab:ranges}
\end{table}

\begin{figure}
    \centering
    \includegraphics[scale=0.35]{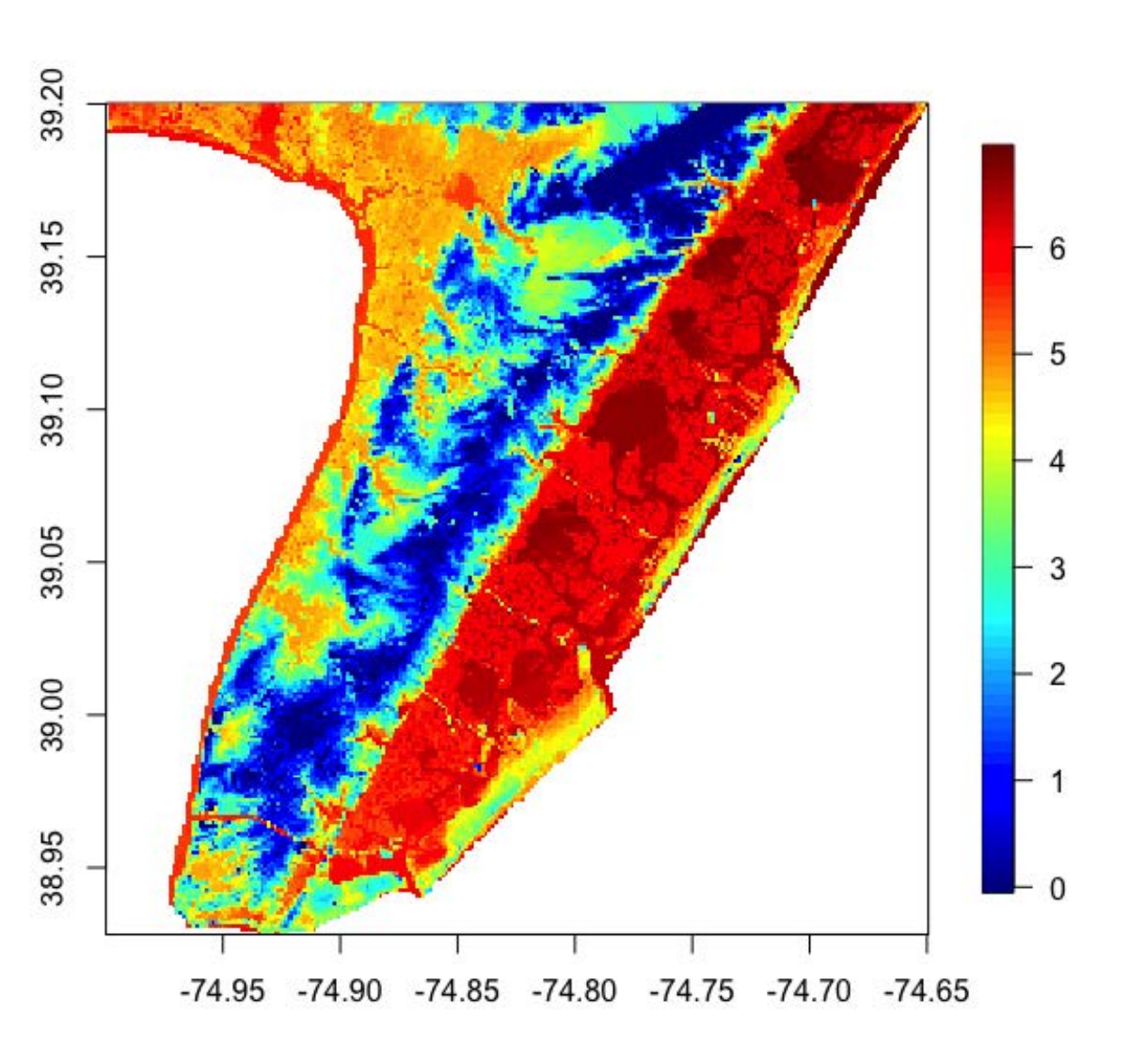}
    \quad 
    \includegraphics[scale=0.6]{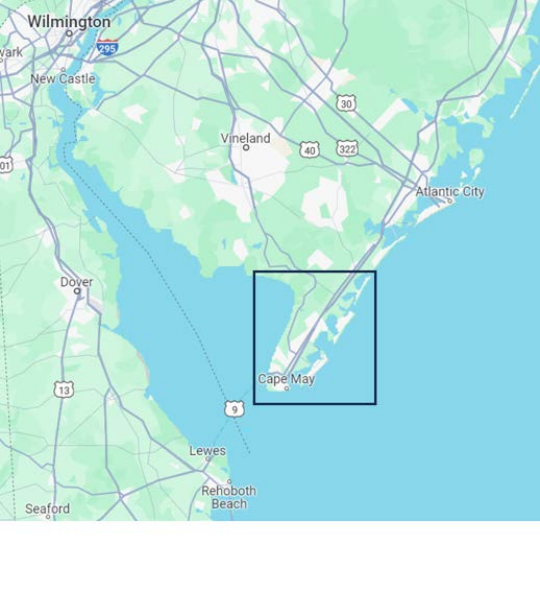}
    \caption{Output from one SLOSH model run, side-by-side with the location of the peninsula in Google Maps. Here, $x$ and $y$ axes represent longitude and latitude, respectively.}
    \label{fig:locations}
\end{figure}

Our SLOSH simulations predict hurricane-induced flooding in the southern tip of New Jersey (see Figure~\ref{fig:locations}). 
%Corresponding to a specific value of the five input parameters, one output from SLOSH is a $4520\times 5115$ rectangular grid of storm surge heights for each of the $23119800$ locations between $-76.11263^0$ and $-70.99763^0$ longitude and $36.71718^0$ and $41.23718^0$ latitude with a spatial resolution of $.001$ degrees in each direction \text. 
Since the major focus is study the damage to electrical power stations, and majority of the power stations are far enough in-land, we only focus on the functional data over $n=49719$ spatial locations. A map of of output from one SLOSH model run is shown in Figure \ref{fig:locations}.

%\underline{\textbf{Local attributes.}} 
In addition, elevation data is available for each spatial location of the electrical power stations. Recognizing the significance of this attribute in assessing the maximum flood height at a given location, we incorporate elevation as a local attribute in our analysis.

%\underline{\textbf{Objective.}} 
Power stations within this region are typically designed to withstand flooding up to a level of \emph{four feet}, with higher levels leading to catastrophic damage. Consequently, our focus is on evaluating the emulator's capability to reliably predict whether a storm surge has exceeded the four-foot mark. This predictive information is crucial for making informed decisions about potential interventions, such as the need to initiate a station shutdown in anticipation of an approaching storm.
%\end{enumerate}

\section{Proposed Approach: A Collaborative Sketching Framework}\label{three_step}
Let $\by_s=(y_s(\bu_1),...,y_s(\bu_n))^T$ and $\bepsilon_s=(\epsilon_s(\bu_1),...,\epsilon_s(\bu_n))^T$ be the vector of responses and errors collected over all index points for the $s$th simulation. Further, assume $\bX$ is an $n\times q$ matrix with its $i$th row given by $\bx(\bu_i)^T$ and $\bw_s=(w_s(\bu_1),...,w_s(\bu_n))^T$ is the vector of functional random effects at all the index points corresponding to the $s$th simulation. The model (\ref{index_model}) yields the Gaussian linear mixed model
\begin{align}\label{eq: stack_model}
\by_s = ({\boldsymbol 1}_n\otimes \bz_s^T)\bgamma + \bX\bbeta + \bw_s+\bepsilon_s\;,\quad \bepsilon_s\sim N(0,\tau^2\bI_n),\:\:s=1,...,S,
\end{align}
where ${\boldsymbol 1}_n$ refers to the $n$-dimensional vector of ones.

Specifying Gaussian process prior on the unknown function $w_s(\bu)\sim GP(0,\sigma^2\kappa(\cdot,\cdot;\theta))$ independently over $s=1,..,S$ leads to $\bw_s\stackrel{i.i.d.}{\sim} N(\bzero, \sigma^2\bK(\theta))$, where 
$\bK(\theta)$ is an $n\times n$ matrix with its $(i,j)$th entry given by $\kappa(\bu_i,\bu_j;\theta)$. Here $\theta$ and $\sigma^2$ are referred to the length-scale parameter and variance parameter, respectively, for the functional regression. Assuming response over simulators are independent, a customary Bayesian hierarchical model is formulated from (\ref{eq: stack_model}) as follows
\begin{align}\label{hier1}
p(\bgamma,\bbeta,\tau^2,\sigma^2)\times \prod_{s=1}^S N(\by_s|({\boldsymbol 1}_n\otimes \bz_s^T)\bgamma + \bX\bbeta, \sigma^2\bK(\theta)+\tau^2\bI_n),
\end{align}
where $p(\bgamma,\bbeta,\tau^2,\sigma^2)$ is the prior distribution on the parameters $(\bgamma,\bbeta,\tau^2,\sigma^2)$. The hierarchical model (\ref{hier1}) fixes the length-scale parameter $\theta$. While the data can inform about this parameter, it is inconsistently estimable for the general Matern class of correlation functions \cite{zhang2004inconsistent} often resulting in poorer convergence. Hence, recent studies proposed an approach where inference is drawn using hierarchical models like (\ref{hier1}) by fitting the model across several fixed values of $\theta$ and subsequently combining the inferences tailored to draw achieve a specific inferential goal \cite{zhang2023exact}. We will adopt a similar strategy of fixing $\theta$ during model fitting with every data sketch. Even with a fixed $\theta$, the Bayesian computation of (\ref{hier1}) involves inverting the $n\times n$ matrix $\sigma^2\bK(\theta)+\tau^2\bI_n$ which is infeasible for the simulator data in Section~\ref{slosh}. 

To circumvent the computational issues posed by $\bK(\theta)$ induced from full GP prior on $w_s$, variants of GP prior are proposed on $w_s$, leading to efficient computation of $\bK(\theta)^{-1}$. This includes low-rank GP priors and sparse GP priors, among others. This article employs modified predictive process (MPP) prior \cite{finley2009hierarchical} and nearest neighbor GP (NNGP) prior \cite{datta2016hierarchical} as the representative low-rank and sparse GP priors, respectively. Let $\tilde{\bu}_1,...,\tilde{\bu}_{n_{knot}}$ be a set of ``knot" points randomly selected from $\mathcal{U}$, with $n_{knot}<<n$. Let $\tilde{\bK}(\theta)$ and $\tilde{\tilde{\bK}}(\theta)$ be $n_{knot}\times n_{knot}$ and $n\times n_{knot}$ matrix with the $(i,j)$th entries given by $\kappa(\tilde{\bu}_i,\tilde{\bu}_j;\theta)$ and $\kappa(\bu_i,\tilde{\bu}_j;\theta)$, respectively. The MPP specifies $\bK(\theta)$ as $\bK(\theta)=\tilde{\tilde{\bK}}(\theta)\tilde{\bK}(\theta)^{-1}\tilde{\tilde{\bK}}(\theta)+\bD(\theta)$, where $\bD(\theta)$ is an $n\times n$ diagonal matrix with the $i$th diagonal entry $1-\tilde{\bkappa}(\bu_i;\theta)^T\tilde{\bK}(\theta)^{-1}\tilde{\bkappa}(\bu_i;\theta)$, $\tilde{\bkappa}(\bu_i;\theta)=(\kappa(\bu_i,\tilde{\bu}_1;\theta),...,\kappa(\bu_i,\tilde{\bu}_{n_{knot}};\theta))^T$. In contrast, NNGP sparsifies $\bK(\theta)^{-1}$ by replacing $w_s(\bu_i)|w_s(\bu_1),...,w_s(\bu_{i-1})$ with $w_s(\bu_i)|\bw_s(\bU_{i,nn})$, where $\bU_{i,nn}$ is the few nearest neighbors of $\bu_i$ \cite{datta2016hierarchical}. While MPP and Sparse GP offers model based efficient alternatives to the full GP,
this article considers an approach orthogonal to them. Specifically, it considers a two-stage distributed Bayesian learning approach, referred to as the collaborative sketching framework, which enables scaling up full-GP, as well as its computationally efficient alternatives (e.g., MPP and Sparse-GP), as elaborated in the subsequent sections.

\subsection{Construction of Sketched Posteriors}\label{sec:Sketched posterior}
At the first stage, we construct multiple low-dimensional random linear mapping or ``sketches" of the full data and fit model (\ref{hier1}) with these data sketches in parallel. To elaborate on this, let $\bPhi_1,...,\bPhi_H$ are $m\times n$ dimensional sketching matrices with random entries encoding random linear mapping of the data to lower dimensions, with $m<<n$. For $h=1,..,H$, the sketching matrix $\bPhi_h$ is applied to $\by_s$, $\bX$ and $\bZ_{s}={\boldsymbol 1}_n\otimes \bz_s^T$ to construct the $m\times 1$ sketched response vector $\by_{s,\bPhi_h} = \bPhi_h \by_s$ and sketched predictor matrices $\bX_{\bPhi_h} = \bPhi_h \bX$ and $\bZ_{s,\bPhi_h} = \bPhi_h \bZ_s$. We will return to the specification of $\bPhi_h$, which, of course, will be crucial for relating the inference from the compressed data with the full model. For now assuming that we have fixed $\bPhi_h$ and a fixed value of the length-scale parameter $\theta_h$, we define the sketched posterior distribution of the model parameters by
\begin{equation}\label{sketched posterior}
\Pi_h(\bgamma,\bbeta,\tau^2,\sigma^2|\bPhi_h,\theta_h)= \frac{p(\bgamma,\bbeta,\tau^2,\sigma^2)\times \prod_{s=1}^S L_{s,h}(\sigma^2,\tau^2,\bgamma,\bbeta)^{n/m}}{\int p(\bgamma,\bbeta,\tau^2,\sigma^2)\times \prod_{s=1}^S L_{s,h}(\sigma^2,\tau^2,\bgamma,\bbeta)^{n/m} d\sigma^2 d\tau^2 d\bgamma d\bbeta},
\end{equation}
where $L_{s,h}(\sigma^2,\tau^2,\bgamma,\bbeta)=N(\by_{s,\bPhi_h}|\bZ_{s,\bPhi_h}\bgamma + \bX_{\bPhi_h}\bbeta, \sigma^2\bPhi_h\bK(\theta_h)\bPhi_h^T+\tau^2\bI_m)$ denotes the likelihood of data sketch from the $s$th simulator run, after fixing $\{\bPhi,\theta\}$ at $\{\bPhi_h,\theta_h\}$. Here $\bK(\theta_h)$ can be induced from full-GP or its computationally efficient variants as discussed before. The likelihood $L_{s,h}$ differs from the one obtained by applying $\mathbf{\Phi}_h$ to the likelihood in (\ref{eq: stack_model}) because the error distribution in $L_{s,h}$ is retained as the usual noise distribution without any influence of $\bPhi_h$. Thus, the likelihood $L_{s,h}$ corresponds to the likelihood derived from a model similar to (\ref{eq: stack_model}), but it is applied to the sketched dataset ${\mathbf{y}_{s,\mathbf{\Phi}_h}, \mathbf{X}_{\mathbf{\Phi}_h}, \mathbf{Z}_{s,\mathbf{\Phi}_h}}$. Employing a $\mathbf{\Phi}_h$-transformed model on (\ref{eq: stack_model}), where noise distribution is transformed by $\mathbf{\Phi}_h\mathbf{\epsilon}_s$
will not deliver the computational benefits.

The modification of likelihood to yield the sketched posterior density in (\ref{sketched posterior}) is referred to as stochastic approximation \cite{guhaniyogi2023distributed,guhaniyogi2022distributed}. Conceptually, raising the likelihood to the power of $n/m$ can be viewed as replicating the sketched data $n/m$ times, so stochastic approximation accounts for the fact that the $h$th sketched posterior distribution $\Pi_h$ conditions on an $m-$dimensional random sketch of the full data and ensures that its variance aligns in the same order (as a function of $n$) as that of the full data posterior distribution.

\subsubsection{Computation of sketched posteriors}\label{computation}
Assume an $IG(a_{\tau},b_{\tau})$, $IG(a_{\sigma},b_{\sigma})$ and $N(\bzero,\bI)$ prior distributions on $\tau^2, \sigma^2$ and $\blambda=(\bbeta^T,\bgamma^T)^T$. 
With the proposed stochastic approximation, the posterior computation of sketched posterior $\Pi_h$ with the $h$th data sketch involves Markov Chain Monte Carlo (MCMC) which cycles through drawing samples from the following distributions:
\begin{itemize}
    \item Sample
    $(\bbeta^T,\bgamma^T)^T|-\sim N(\bmu_{\bbeta,\bgamma}, \; \bSigma_{\bbeta,\bgamma})$, where 
    $\bSigma_{\bbeta,\bgamma} = \left\{(m/n) \sum_{s=1}^{S} \mathbf{A}_s^T \bSigma^{-1} \mathbf{A}_s + \mathbf{I}_p \right\}^{-1}$,  $\bmu_{\bbeta,\bgamma} = \bSigma_{\bbeta,\bgamma} \left\{ (m/n)\sum_{s=1}^{S} \mathbf{A}_s^T\bSigma^{-1} \by_{s,\bPhi_h} \right\}$. Here $\bA_s=[\bX_{\bPhi_h}:\bZ_{s,\bPhi_h}]$ is an $m\times n$ matrix and 
    $\bSigma=(\bPhi_h\bK(\theta_h)\bPhi_h^T+\tau^2\bI_m)$ is an $m\times m$ matrix. This step incurs a computation complexity of $O(m^3)$, since $\bSigma$ is an $m \times m$ covariance matrix that needs to be inverted.
\item Sample $\sigma^2$ and $\tau^2$ through Metropolis-Hasting sampler. Since $\theta_h$ is kept fixed throughout the analysis of $\Pi_h$, we need to compute $\boldsymbol{\Phi}^\intercal \bK(\theta_h) \boldsymbol{\Phi}$ only once which leads to substantial computational benefit. Additionally, the strategy needs storage of the $m\times m$ matrix $\boldsymbol{\Phi}^\intercal \bK(\theta_h) \boldsymbol{\Phi}$ instead of the full data covariance matrix, which reduces the storage cost from $O(n^2)$ to $O(m^2)$.
\end{itemize}
Next we focus on drawing predictive inference at $n^*$ index points $\mathcal{U}=\{\bu_1^*,...,\bu_{n^*}^*\}$ for a new simulator run. To this end, let $\by^*$ and $\bX^*$ be $n^*\times 1$ dimensional response vector and $n^*\times q$ dimensional matrix of local attributes at $n^*$ index points, respectively. Assume $\bz^*$ is the input attributes corresponding to the new simulator run, $\bZ^*=\mathbf{1}_{n^*}\otimes \bz^{*T}$, and $\bw_s^*=(w_s(\bu_1^*),...,w_s(\bu_{n^*}^*))^T$. The posterior predictive density of $\by^*$ is given by
\begin{align}
\begin{split}\label{posterior_predictive}
& p(\by^*|\bX^*,\bZ^*,\bX_{\bPhi_h}, \bZ_{s,\bPhi_h}, \by_{s,\bPhi_h})=
\int f(\by^*|\bX^*,\bZ^*,\bgamma,\bbeta,\sigma^2,\tau^2)\Pi_h(\bgamma,\bbeta,\sigma^2,\tau^2|\bPhi_h,\theta_h)\\
& f(\by^*|\bX^*,\bZ^*,\bgamma,\bbeta,\sigma^2,\tau^2) = N(\by^*|\bmu^*+\bSigma_{21}\bSigma_{11}^{-1}(\by-\bmu),\bSigma_{22}-\bSigma_{21}\bSigma_{11}^{-1}\bSigma_{12})\\
& \bmu^*=(\mathbf{1}_{n^*}\otimes \bz^*)\bgamma+\bX^*\bbeta, \:\bmu=(\mathbf{1}_{n}\otimes \bz)\bgamma+\bX\bbeta\\
&\bSigma_{22}=\text{Var}(\bw_s^*), \:\bSigma_{11}=\text{Var}(\bw_s), \:\bSigma_{12}=\bSigma_{21}^T=\text{Cov}(\bw_s, \bw_s^*).
\end{split}
\end{align}
We utilize composition sampling to generate MCMC samples from the posterior predictive density (\ref{posterior_predictive}) of $\by^*$. The entire analysis of $\Pi_1,...,\Pi_H$ will be distributed across $H$ separate CPUs, enabling parallel processing for enhanced efficiency. 
%, and is somewhat detrimental to the cause of data confidentiality (as in that case, the analyst need to know $\bPhi$) that are provided by (\ref{eq: svc_basic_compressed_bhm})
\subsubsection{Choice of Sketching Matrices}
To specify the matrix $\bPhi_h$, we adopt the concept of data oblivious Gaussian sketching as proposed in \cite{sarlos2006improved}. This involves independently selecting elements of $\bPhi_h$ from a normal distribution with mean 0 and variance $1/n$, and then keeping them fixed. The dominant computational effort required for generating the sketched data using Gaussian sketches follows a time complexity of $O(mn^2)$. While there are alternative data oblivious methods available for efficiently sketching $\bPhi_h$, like the Hadamard sketch \cite{ailon2009fast} and the Clarkson-Woodruff sketch \cite{clarkson2017low}, these options hold less relevance in Bayesian contexts. This is due to the fact that the computational time needed for the operation in (\ref{sketched posterior}) is significantly greater than that needed for constructing the sketching matrix itself.
The sketched data serves as a surrogate for the Bayesian estimation of model coefficients. Given that the count of sketched records is much smaller than the total records in the complete data matrix, the process of fitting the model becomes both computationally efficient and resource-friendly. This efficiency extends to storage requirements as well as the count of floating-point operations (flops) needed, as outlined in Section~\ref{computation}.

\subsubsection{Data Privacy}\label{sec:privacy}
Crucially, even when $\bPhi_h$ is known, the linear systems $\bPhi_h\by_s$, $\bPhi_h\bX$, and $\bPhi_h\bZ_s$ are significantly under-determined due to the substantial difference in the dimensions $m$ and $n$. This results in a safeguarding of the privacy of the data samples. To evaluate the privacy implications in terms of information theory, an upper bound on the average mutual information per unit can be used, denoted as $I(\bX_{\bPhi_h},\bX)/nq$. It can be shown that the supremum of $I(\bX_{\bPhi_h},\bX)/nq$ is bounded by $O(m/n)$ \cite{zhou2008compressed}, where the supremum is taken across all possible distributions of $\bX$.
As $m$ grows at a much slower rate than $n$, and $n\rightarrow\infty$, the supremum of the average mutual information approaches 0. This implies that, intuitively, the compressed data reveals no more information about the original data than what could be derived from an independent sample. It's important to note that this bound is derived under the assumption that $\bPhi_h$ is known. In practical scenarios, only $\bX_{\bPhi_h}$ (along with $\by_{s,\bPhi_h}$ and $\bZ_{s,\bPhi_h}$) would be disclosed to the analyst, without revealing the actual matrix $\bPhi_h$ itself. Consequently, the privacy imposed through sketching turns out to be more stringent than what is implicated by this theoretical outcome. It is worth emphasizing that these privacy implications stem from the information theory and are distinct from the broader notion of data privacy explored through concepts like $\epsilon$-differential privacy.

\subsubsection{Distributed Storage in Research Centers} \label{subsec:storage}
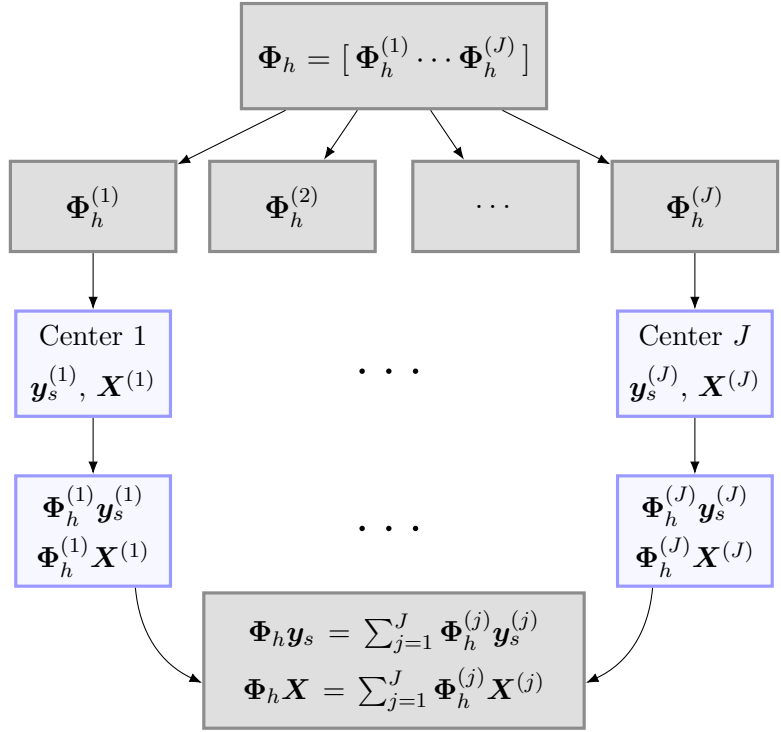
\begin{figure}[tb]
\begin{center}
    \begin{tikzpicture}[
            rect/.style={rectangle, draw=blue!40, fill=blue!3, very thick, minimum width=18mm, minimum height=14mm, text width=18mm, align=center, execute at begin node=\setlength{\baselineskip}{1.75em}},
            long_rect/.style={rectangle, draw=gray!90, fill=gray!25, very thick, minimum width=18mm, minimum height=14mm, text width=38mm, align=center},
            text_rect/.style={rectangle, draw=gray!90, fill=gray!25, very thick, minimum width=22mm, minimum height=12mm, align=center},
            end_rect/.style={rectangle, draw=gray!90, fill=gray!25, very thick, minimum width=22mm, minimum height=18mm, align=center, text width=48mm, execute at begin node=\setlength{\baselineskip}{2em}},
            steps/.style={text width=3.25cm},
            steps_big/.style={text width=5.25cm},
            ]
        
        \node[long_rect] (start) at (1,0) {\large $\bPhi_h = [\, \bPhi_h^{(1)} \cdots \bPhi_h^{(J)} \, ] $};
        \node[text_rect] (phi_1) at (-3,-2) {\large $\bPhi_h^{(1)}$};
        \node[text_rect] (phi_2) at (-0.35,-2) {\large $\bPhi_h^{(2)}$};
        \node[text_rect] (phi_mid) at (2.35,-2) {\large \ldots};
        \node[text_rect] (phi_J) at (5,-2) {\large $\bPhi_h^{(J)}$};
        %\node[steps] (step0) [left=4.5cm of start] {Full Data: };

        \node[rect] (center1) [below=.75cm of phi_1] {Center 1 $\by_s^{(1)}$, $\bX^{(1)}$};  
        \node[rect] (centerJ) [below=.75cm of phi_J] {Center $J$ $\by_s^{(J)}$, $\bX^{(J)}$};
        \node[] (center_mid) [below=3.3cm of start] {\Huge \ldots};
        \node[] (data_mid) [below=1.75cm of center_mid] {\Huge \ldots};
        
        \node[rect] (data1) [below=.75cm of center1] {$\bPhi_h^{(1)}\by_s^{(1)}$  $\bPhi_h^{(1)}\bX^{(1)}$};
        \node[rect] (dataJ) [below=.75cm of centerJ] {$\bPhi_h^{(J)}\by_s^{(J)}$  $\bPhi_h^{(J)}\bX^{(J)}$};

        \node[end_rect] (end) [below = 6.4cm of start] {$\bPhi_h\by_s = \sum_{j=1}^J \bPhi_h^{(j)}\by_s^{(j)}$ $\bPhi_h\bX = \sum_{j=1}^J \bPhi_h^{(j)}\bX^{(j)}$};

        \path (start) edge (phi_1);
        \path (start) edge (phi_2);
        \path (start) edge (phi_mid);
        \path (start) edge (phi_J);
        \path (phi_1) edge (center1);
        \path (phi_J) edge (centerJ);
        \path (center1) edge (data1);
        \path (centerJ) edge (dataJ);
        \path (data1) edge [bend right=30] (end);
        \path (dataJ) edge [bend left=30] (end);
        %\path(sketched1) edge node[right,pos=.5] {\large $\theta_1$} (posterior1);
        
        %%%%%%%%%%%%%%%%%%%%%%%%%%%%%%%%%%%%%%%%%%%%%%%%%%%%%%%%%%%

    \end{tikzpicture}
    \caption{Distributed storage of data for $J$ separate storage centers.}
    \label{fig:dist_storage}
\end{center}
\end{figure}

An essential benefit of the proposed framework lies in its capacity to analyze data collectively from research centers that are restricted from sharing their data with one another. Consider a scenario with $J$ centers denoted as $Center_1, ..., Center_J$, where the $j$th center holds data for $n_j$ indexing points, $n = \sum_{j=1}^J n_j$. In this context, the $j$th center privately stores an $n_j$-dimensional response vector $\by_s^{(j)}$ and an $n_j \times p$-dimensional matrix $\bX^{(j)}$ for local attributes, such that $\by_s = (\by_s^{(1)}, ..., \by_s^{(J)})^T$ and $\bX = [\bX^{(1)T} : \cdots : \bX^{(J)T}]^T$. Let $\bPhi_h = [\bPhi_h^{(1)T} : \cdots : \bPhi_h^{(J)T}]^T$, where $\bPhi_h^{(j)}$ represents the $m \times n_j$ sketching matrix. The sketching matrix $\bPhi_h^{(j)}$ can be constructed within each research center, and local computations of $\bPhi_h^{(j)}\by_s^{(j)}$ and $\bPhi_h^{(j)}\bX^{(j)}$ can be performed within the $j$th center before releasing them to calculate the $h$th sketched posterior. The privacy guarantee, as discussed in Section~\ref{sec:privacy}, ensures that the original data cannot be reconstructed from the sketches obtained from centers. By aggregating these local sketches $\bPhi_h\by_s=\sum_{j=1}^J \bPhi_h^{(j)}\by_s^{(j)}$ and $\bPhi_h\bX=\sum_{j=1}^J\bPhi_h^{(j)}\bX^{(j)}$, the framework allows the computation of complete data sketches. The schematic representation in the Figure~\ref{fig:dist_storage} illustrates the strategy for securely computing sketched posteriors while preserving privacy. 

\subsection{Construction of the Collaborative Sketched Posterior}
The approach described below for combining sketched posteriors constructed in Section~\ref{sec:Sketched posterior} is in the same spirit as the combination of ``subset posteriors" outlined in the divide-and-conquer strategies in Bayesian inference \cite{guhaniyogi2023distributed}. The salient feature of the combination technique is that it is agnostic to the model- or data-specific assumptions, such as assuming independence among samples in the training data. 

\begin{figure}[tb]
%\begin{center}
    \begin{tikzpicture}[
            rect/.style={rectangle, draw=blue!40, fill=blue!3, very thick, minimum width=22mm, minimum height=10mm},
            text_rect/.style={rectangle, draw=gray!90, fill=gray!25, very thick, minimum width=22mm, minimum height=12mm, align=center, text width=20mm},
            sketched/.style={rectangle, draw=gray!90, fill=gray!25, very thick, minimum width=3cm, minimum height = 8mm},
            empty
            intersect/.style={circle, draw=black, fill=black, inner sep=0pt, minimum size=4mm},
            steps/.style={text width=3.25cm},
            steps_big/.style={text width=5.5cm},
            ]
        
        \node[rect] (start) at (1,0) {\large $\by_s$, $\bZ_s$, $\bX$};
        \node[steps] (step0) [left=4.35cm of start] {Full Data: };

        \node[sketched] (sketched1) at (-2, -2) {$\bPhi_1 \by_s$, $\bPhi_1 \bX$};
        \node[] (sketch_dots) [below=1.3cm of start] {\Huge \ldots};
        \node[sketched] (sketchedH) at (4, -2) {$\bPhi_H \by_s$, $\bPhi_H \bX$};
        \node[steps] (step1) [left=.975cm of sketched1] {Sketched Data: };
        
        \path (start) edge node[left,pos=0.4] {\large $\bPhi_1$} (sketched1);
        \path (start) edge node[right,pos=0.4] {\large $\bPhi_H$} (sketchedH);

        \node[sketched] (posterior1) [below=1.5cm of sketched1] {$\pi_1(\bgamma, \bbeta, \tau^2, \sigma^2 \, | \, \bPhi_1, \theta_1)$};
        \node[] (post_dots) [below=3.65cm of start] {\Huge \ldots};
        \node[sketched] (posteriorH) [below=1.5cm of sketchedH] {$\pi_H(\bgamma, \bbeta, \tau^2, \sigma^2 \, | \, \bPhi_H, \theta_H)$};
        \node[steps] (step2) [left=0.52cm of posterior1] {Sketched Posteriors: };

        \path(sketched1) edge node[right,pos=.5] {\large $\theta_1$} (posterior1);
        \path(sketchedH) edge node[right,pos=.5] {\large $\theta_H$} (posteriorH);

        \node[rect] (end) [below=5.5cm of start] {\large $\Bar\pi = \underset{\pi \in P(\Omega)}{\text{argmin }} \frac{1}{H}\sum_{h=1}^H W_2^2(\pi, \pi_h)$};
        \node[steps_big] (step3) [left=0.4cm of end] {Collaborative Sketched Posterior: };

        \path (posterior1) edge (end);
        \path (posteriorH) edge (end);
        
        %%%%%%%%%%%%%%%%%%%%%%%%%%%%%%%%%%%%%%%%%%%%%%%%%%%%%%%%%%%

    \end{tikzpicture}
    \caption{Visual overview of distributed learning algorithm for $H$ parallel nodes.}
    \label{fig:algorithm_viz}
%\end{center}
\end{figure}
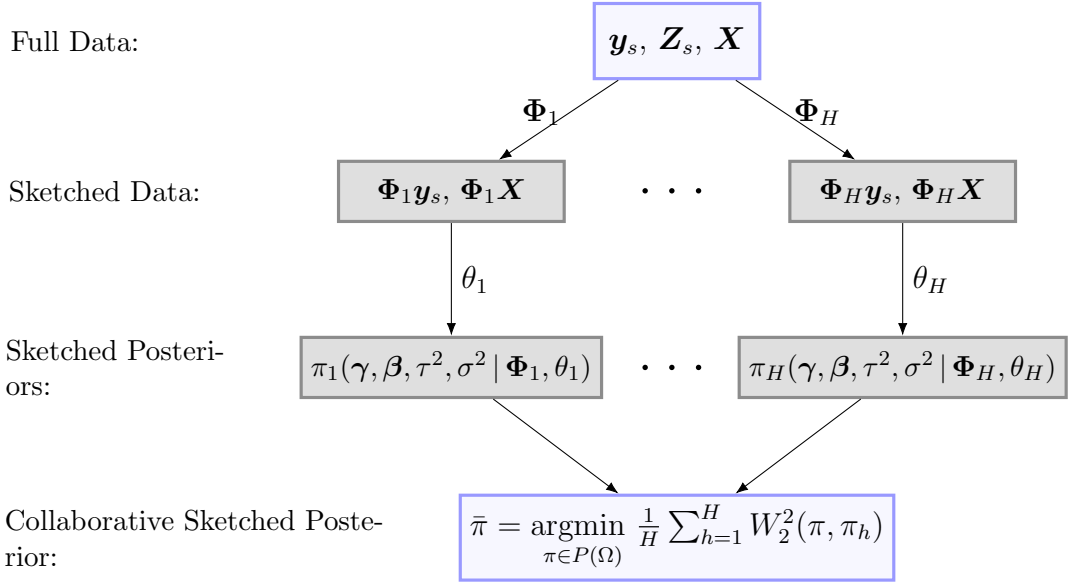

This process of combining sketched posteriors involves leveraging the concept of the Wasserstein barycenter, a technique extensively used in the realm of scalable Bayesian methods for distributed inference, as highlighted in works such as \cite{guhaniyogi2023distributed} and \cite{srivastava2018scalable}. Let $(\Omega,\rho)$ be a complete separable metric space and $\mathcal{P}_2(\Omega)$ denotes the probability distributions on $(\Omega,\rho)$ with finite second moments. Let $\tilde{\Pi}_1,\tilde{\Pi}_2$ be two probability measures in $\mathcal{P}_2(\Omega)$. Assume $\Delta (\tilde{\Pi}_1, \tilde{\Pi}_2)$ is the set of all probability measures on $\Omega \times \Omega$ with marginals $\tilde{\Pi}_1$, $\tilde{\Pi}_2\in \mathcal{P}_2(\Omega)$. Then the Wasserstein distance of order 2, denoted as $W_2$, between $\tilde{\Pi}_1,\tilde{\Pi}_2$ is defined as
$W_2(\tilde{\Pi}_1, \tilde{\Pi}_2) = \{\underset{\pi \in \Delta (\tilde{\Pi}_1, \tilde{\Pi}_2)} {\mathrm{inf}} \, \int_{\Omega \times \Omega} \rho^2(x, y) \, d \pi(x, y) \}^{1/2}.$ In the context of Section~\ref{sec:Sketched posterior}, if $\Pi_1, \ldots, \Pi_H$ all have finite second moments, then the Wasserstein barycenter of $\Pi_1, \ldots, \Pi_H$ is defined as
\begin{align}
  \bar{\Pi} = \underset{ \Pi \in \mathcal{P}_2(\Omega)} {argmin} \frac{1}{H} \sum_{h=1}^{H} W_2^2 (\Pi, \Pi_h). \label{eqn:w-bary}
\end{align}
It is known that $\bar{\Pi}$ exists and is unique \cite{agueh2011barycenters}. The Wasserstein barycenter $\bar{\Pi}$ is referred to as the collaborative sketched posterior for the parameters and replaces full data posterior as its computationally efficient approximation for inference on (\ref{eq: stack_model}). The two-step procedure for the construction of collaborative sketched posterior can be visualized in Figure \ref{fig:algorithm_viz}.

When focus lies on drawing inference on one-dimensional functional of the parameters, denoted as $\alpha$, the sketched pseudo posterior distribution of $\alpha$ can be easily derived by averaging the empirical sketched posterior quantiles \cite{guhaniyogi2023distributed}. This is facilitated by the fact that the Wasserstein distance ($W_2$ distance) between two univariate distributions corresponds to the $L_2$ distance between their respective quantile functions (as demonstrated in Lemma 8.2 of \cite{bickel1981some}). More precisely,
for a given quantile level $\xi \in (0,1)$, let $\hat{\alpha}_{\xi,h}$ represent the $\xi$th empirical quantile of $\alpha$ based on MCMC samples drawn from the marginal posterior distribution of $\alpha$ from $h$th subset.
With $\hat{\alpha}_{\xi}$ as the $\xi$th empirical quantile of $\alpha$ following the collaborative sketched posterior, $\hat{\alpha}_{\xi}$ can be expressed as,
\begin{align}\label{xi_grid}
\hat{\alpha}_{\xi}=\frac{1}{H}\sum_{h=1}^H \hat{\alpha}_{\xi,h},
\end{align}
where the value of $\xi$ is varied across a fine grid within the interval (0,1). By utilizing a sufficiently fine $\xi$-grid in the equation (\ref{xi_grid}), MCMC samples from the marginal collaborative sketched posterior distribution of $\alpha$ can be obtained by inverting the empirical distribution function supported on the estimated quantiles.

In real-world applications, the primary focus often centers on the marginal distributions of model parameters and predicted values, i.e., on one-dimensional functionals of parameters. Consequently, the process of obtaining the univariate Wasserstein barycenter by averaging quantiles, as described in equation (\ref{xi_grid}), accomplishes this objective with remarkable versatility and a user-friendly implementation. Due to its general applicability and ease of use, in the subsequent sections, we concentrate exclusively on scenarios where $\alpha$ is one-dimensional and utilize equation (\ref{xi_grid}) to calculate the collaborative sketched posterior by utilizing its empirical quantiles. It is worth noting that there are existing approaches which allow combination of joint posteriors instead of marginal posteriors, see the recent work by \cite{guhaniyogi2022distributed} and the references therein. However, these methods are computationally more demanding and yield only marginal improvements over the univariate quantile combination approach mentioned in equation (\ref{xi_grid}).

While our proposed approach shares similarities with the evolving field of distributed Bayesian inference for correlated data \cite{guhaniyogi2023distributed,guhaniyogi2022distributed,guhaniyogi2018meta}, it offers a distinct advantage over these approaches. Notably, all the existing distributed Bayesian learning approaches involve dividing the data into shards based on a user-defined partitioning scheme. However, a crucial observation from simulation studies in \cite{guhaniyogi2023distributed} is that the effectiveness of distributed inference is somewhat influenced by the specific choice of data shards. This necessitates application-specific judgments when partitioning data into these shards. In contrast, our approach sidesteps the need to create data shards and instead focuses on constructing data sketches with multiple random sketching matrices. The fundamental distinction lies in the fact that while the Wasserstein barycenter of shard posteriors in traditional distributed learning methods is contingent upon the chosen data partitioning, our approach incorporate data sketching as an intrinsic facet of the model development process, and mitigates the impact of variations due to the choice of sketching matrices by computing the Wasserstein barycenter of sketched posteriors, each constructed using a different random sketching matrix. This innovation leads to the novel concept of \emph{robust distributed Bayesian learning}, where the resulting inference is more resilient to the uncertainties introduced by random sketching. 

%\section{Theoretical Performance}\label{theory}
\section{Simulation Studies}\label{simulations}
One salient feature of the proposed approach is that it 
is agnostic to the dimension $d$ of the index points. However, 
we empirically validate the efficacy of the collaborative sketched posterior for the case of $d=2$, where the index points represent locations in space, since $d=2$ for the SLOSH emulator data. 
To comprehensively evaluate the performance of the collaborative sketched posterior, it is constructed with $w_s(\bu)$ modeled using each of the three strategies: (i) the full-rank GP, (ii) the low-rank Modified predictive process \cite{finley2009hierarchical} and (iii) NNGP \cite{datta2016hierarchical}. The inferential and predictive performance of the collaborative sketched posterior under (i)-(iii) is compared with existing distributed Bayesian approaches utilizing different data subsetting techniques. To simulate the data, we use $n+n_0$
spatial locations $\bu_1,\ldots,\bu_{n+n_0}$ drawn uniformly over the domain
$\mathcal{D} = [0,10]\times [0,10]$. The data generation procedure involves 
$p=1$ global attribute and $q=2$ local attributes. The global attribute values are generated from the $N(0,1)$ distribution for $S+S_0$ simulations, i.e., $z_1,...,z_{S+S_0}\stackrel{i.i.d.}{\sim} N(0,1)$. Similarly, the local attributes $\bx(\bu_1),...,\bx(\bu_{n+n_0})$ are simulated independently from $N(\bzero,\bI)$. For each $s=1,...,S+S_0$ and $i=1,...,n+n_0$, the response 
$y_s(\bu_i)$ is drawn independently from $N(z_s\gamma_0+\bx(\bu_i)^T\bbeta_0+w_{s}(\bu_i),\tau_0^{2})$ following (\ref{index_model}), where the noise variance $\tau_0^{2}$ is set to be $0.2$.

The true coefficient $\gamma_0$ for the global attribute is set to be $5$, where as the true coefficient $\bbeta_0$ for local attributes is kept at $(2,-1)$. The true function $w_{s0}(\bu)$s are generated from a Gaussian process with mean $0$ and covariance kernel $\sigma_0^{2}\kappa(\cdot,\cdot;\theta_0)$ independently over $s$, i.e., $(w_{s0}(\bu_1),...,w_{s0}(\bu_{n+n_0}))^{\T}\stackrel{ind.}{
\sim} N(\bzero,\sigma_0^{2}\bK(\theta_0))$ for $s=1,...,S+S_0$, where $\bK(\theta_0)$ is an $(n+n_0)\times (n+n_0)$ matrix with the $(j,j')$th element $\kappa(\bu_j,\bu_{j'};\theta_0)$. We set the covariance kernel $\kappa(\cdot,\cdot;\theta_0)$ to be the exponential correlation function given by $\kappa(\bu_j,\bu_{j'};\theta_0) = \exp\left(-\theta_0||\bu_j-\bu_{j'}||\right)$, with the true values $\sigma_0^{2}$ and $\theta_0$ set to $2$ and $3$, respectively. 

Out of $S+S_0=15$ simulations, we randomly choose $S=10$ simulations for model fitting, and the remaining $S_0=5$ simulations are used for testing the predictive performance of the model. For the $S=10$ training simulations, the model is fitted on $n=10,000$ spatial locations, whereas prediction is performed on $n_0=1000$ different locations for the remaining $S_0=5$ simulations. In other words, we assess predictive accuracy not only on different locations in space but also on entirely different simulations.

Throughout all simulations, the collaborative sketched posterior is calculated using a sketching dimension of $m=500$. Moreover, when fitting sketched posteriors with the MPP for each $w_s(\bu)$, we employ $500$ knots, and for fitting NNGP for each $w_s(\bu)$, we consider $10$ nearest neighbors.

\subsection{Competitors}
For all simulations, we compare the collaborative sketching framework proposed here with the Distributed Kriging (DISK) approach \cite{guhaniyogi2023distributed}, a divide-and-conquer Bayesian approach that computes subset posteriors with user-defined data subsets and combines marginal distributions of these subset posteriors by averaging their quantiles, similar to the strategy in (\ref{xi_grid}). To demonstrate sensitivity in inference due to the choice of different data subsetting, DISK is fitted with two different strategies: (a) subdomain and (b) stratification. The subdomain strategy divides the domain into rectangular subdomains with each subset constituting all data points from a subdomain, while the stratification strategy constructs each subset with representative samples from each subdomain. We refer to them as DISK-subdomain and DISK-stratified, respectively, and they both maintain the number of samples per subset to be around the sketching dimension $m=500$ to ensure a fair comparison with our collaborative sketching approach. %Additionally, we compare our approach with BASS \cite{francom2020bass}, a commonly used approach in national laboratories for analyzing functional data from multiple simulations. BASS is not able to incorporate local attributes (i.e., elevation in the SLOSH data) and is susceptible to high variance of estimation unless the number of simulations $S$ is high. However, given its popularity among researchers in national laboratories for the functional data analysis, we include it as a competitor and implement it using the \texttt{R} package \texttt{BASS}.

\subsection{Metrics of Comparison}
Let $\bY^{(l)}$ denote the $l$th post burn-in MCMC sample from the posterior predictive distribution of $\bY=(y_s(\bu_i): i=n+1,...,n+n_0;\:s=S+1,...,S+S_0)^T$, for $l=1,..,L$.
The point prediction for competitors is assessed using mean squared prediction error (MSPE), defined as the discrepancy between the true and the predicted responses $||\bar{\bY}-\bY_t||^2/(n_0S_0)$, where $\bar{\bY}=\frac{1}{L}\sum_{l=1}^L \bY^{(l)}$ and $\bY_t$ denotes the true value of $\bY$.
%$\hat{y}_s(\bu_i)$ be the predicted response at the $i$th out of the sample location for $s$th out of the sample simulation, for $i=n+1,...,n+n_0$ and $s=S+1,...,S+S_0$. The point prediction for competitors is assessed using mean squared prediction error (MSPE), defined as the discrepancy between the true and the predicted responses $\sum_{s=S+1}^{S+S_0}\sum_{i=n+1}^{n+n_0}(\hat{y}_s(\bu_i)-y_s(\bu_i))^2/(n_0S_0)$. 
The predictive uncertainty is determined with the coverage probability, interval score and energy score \cite{gneiting2007strictly} for 95\% predictive intervals (PIs) for all the competing methods over $n_0$ out of sample locations in $S_0$ out of sample simulations. Interval scores favors model with the smallest possible intervals that still contain the data. On the other hand, energy score is a multivariate extension to Continuous Rank Probability Score (CRPS) and is calculated as $\frac{1}{L}\sum_{l=1}^L||\bY^{(l)}-\bY_t||-\frac{1}{2L^2}\sum_{l=1}^L\sum_{l'=1}^L||\bY^{(l)}-\bY^{(l')}||$. Energy score takes into account not only the predictive accuracy of each sample from the posterior predictive distribution, but also the level of uncertainty in the distribution. For this reason, energy score has gained interest in recent literature as a model ranking mechanism \cite{heaton2019case}. Finally, we compare performance of all distributed Bayesian methods for parameter estimation using the posterior medians or point estimates and the 95\% credible intervals (CIs) for
$\bbeta=(\beta_1,\beta_2),\gamma,\sigma^2,\tau^2$. 

\subsection{Results of Simulated Data Analysis}
 
\begin{table}[h]
    \tiny
    \centering
    \begin{tabular}{rrccccc}
    \toprule[1.5pt]
         Model & Competitors & $\sigma^2$ & $\tau^2$ & $\beta_1$ & $\beta_2$ & $\gamma$ \\
         \midrule
        \multicolumn{2}{c}{\textbf{Truth}} & 2.00 & 0.20 & 2.00 & -1.00 & 5.00 \\
        \midrule
Full GP & DISK-Subdomain & 2.13 (2.09, 2.16) & 0.21 (0.21, 0.22) & 1.77 (1.39, 2.14) & -1.00 (-1.01, -0.98) & 5.00 (4.99, 5.01) \\
        & DISK-Stratified & 1.94 (1.91, 1.96) & 0.20 (0.19, 0.21) & 1.99 (1.89, 2.08) & -1.02 (-1.02, -0.98) & 5.00 (5.00, 5.00) \\
        %& Multiplets & 2.10 (2.07, 2.13) & 0.21 (0.20, 0.22) & 2.01 (1.94, 2.08) & -1.01 (-1.02, -0.97) & 5.00 (5.00, 5.00) \\
        %& Random     & 1.98 (1.95, 2.01) & 0.21 (0.20, 0.22) & 2.00 (1.93, 2.07) & -1.00 (-1.01, -0.96) & 5.00 (5.00, 5.00) \\
        & Collaborative Sketching  & 2.07 (2.03, 2.10) & 0.21 (0.19, 0.22) & 2.00 (1.98, 2.01) & -0.99 (-1.00, -0.97) & 5.00 (5.00, 5.00) \\
        \midrule
    MPP & DISK-Subdomain & 1.62 (1.59, 1.65) & 0.18 (0.17, 0.20) & 1.79 (1.55, 2.03) & -1.00 (-1.02, -0.98) & 5.00 (4.99, 5.00) \\
        & DISK-Stratified & 1.80 (1.79, 1.82) & 0.05 (0.04, 0.05) & 1.94 (1.86, 2.01) & -0.99 (-1.02, -0.97) & 5.00 (5.00, 5.00) \\
        %& Multiplets & 2.08 (2.05, 2.11) & 0.08 (0.07, 0.10) & 1.98 (1.91, 2.05) & -0.99 (-1.01, -0.97) & 5.00 (4.99, 5.00) \\
        %& Random     & 1.99 (1.97, 2.01) & 0.03 (0.03, 0.04) & 1.98 (1.92, 2.03) & -0.99 (-1.01, -0.96) & 5.00 (4.99, 5.00) \\
        & Collaborative Sketching  & 2.07 (1.99, 2.16) & 0.11 (0.10, 0.13) & 2.01 (2.00, 2.03) & -0.95 (-0.96, -0.95) & 5.00 (5.00, 5.00) \\
        \midrule
NNGP & DISK-Subdomain & 2.14 (2.10, 2.18) & 0.21 (0.21, 0.22) & 1.76 (1.37, 2.14) & -1.00 (-1.01, -0.98) & 5.00 (5.00, 5.01) \\
        & DISK-Stratified & 1.99 (1.97, 2.01) & 0.21 (0.20, 0.21) & 1.99 (1.90, 2.08) & -1.00 (-1.01, -0.98) & 5.00 (4.99, 5.00) \\
        %& Multiplets & 2.08 (2.05, 2.11) & 0.22 (0.20, 0.22) & 2.01 (1.93, 2.08) & -0.99 (-1.01, -0.97) & 5.00 (5.00, 5.00) \\
       % & Random     & 1.98 (1.96, 2.01) & 0.20 (0.19, 0.22) & 2.00 (1.94, 2.07) & -0.98 (-1.01, -0.96) & 5.00 (5.00, 5.00) \\
        & Collaborative Sketching  & 2.04 (2.00, 2.07) & 0.19 (0.18, 0.20) & 1.97 (1.95, 1.99) & -0.98 (-0.99, -0.97) & 5.00 (5.00, 5.00) \\
        \bottomrule[1.5pt]
    \end{tabular}
    \caption{We calculate the posterior median with 95\% confidence intervals for all model parameters for all the distributed Bayesian competitors, fitting a full-GP, low-rank modified predictive process (MPP), and NNGP. The MPP utilizes 500 knots for model implementation and NGGP uses $10$ nearest neighbors. We set both the sketching dimension for our approach and the size of each subset for the DISK approach to be $m=500$ to ensure comparability.}
    \label{tab:sim_results}
\end{table}

%\begin{figure}[h]
%    \centering
%    \includegraphics[scale=0.10]{figures/pred_diagnostics.png}
%    \caption{MSPE and interval score for the same simulations displayed in Table \ref{tab:sim_results}.}
%    \label{fig:sim_pred_diagnostics}
%\end{figure}

\begin{table}[h]
    \scriptsize
    \centering
    \begin{tabular}{rrcccc}
    \toprule[1.5pt]
         Model & Competitors & MSPE & Coverage & Interval Score & Energy Score\\
        \midrule
Full GP & DISK-Subdomain & 2.44 & 0.95 & 7.02 & 0.89 \\
        & DISK-Stratified & 2.30 & 0.94 & 6.89 & 0.86 \\
        %& Multiplets & 2.30 & 0.95 & 6.87 \\
        %& Random     & 2.30 & 0.94 & 6.88 \\
        & Collaborative Sketching  & 2.27 & 0.95 & 6.87 & 0.86 \\
        \midrule
    MPP & DISK-Subdomain & 2.50 & 0.91 & 7.24 & 0.90 \\
        & DISK-Stratified & 2.32 & 0.92 & 7.05 & 0.89 \\
        %& Multiplets & 2.31 & 0.94 & 6.90  \\
        %& Random     & 2.31 & 0.93 & 6.94 \\
        & Collaborative Sketching  & 2.30 & 0.95 & 6.87 & 0.86 \\
        \midrule
NNGP & DISK-Subdomains & 2.43 & 0.95 & 7.00 & 0.89 \\
        & DISK-Stratified & 2.29 & 0.94 & 6.88 & 0.86 \\
        %& Multiplets & 2.30 & 0.95 & 6.87 \\
        %& Random     & 2.30 & 0.94 & 6.88 \\
        & Collaborative Sketching  & 2.28 & 0.95 & 6.88 & 0.86 \\
        \bottomrule[1.5pt]
    \end{tabular}
    \caption{MSPE, coverage, interval score and energy score for all competing methods.}
    \label{tab:sim_pred_diagnostics}
\end{table}
The simulation results presented in Table~\ref{tab:sim_results} highlight significant sensitivity in parameter estimation, contingent upon the nature of data partitioning in the DISK approach. For instance, when a full GP or NNGP is fitted to each data subset, the 95\% CIs for $\sigma^2$ in DISK-Subdomain and DISK-Stratified do not overlap. Additionally, the 95\% CI for $\beta_1$ exhibits a much larger confidence interval for DISK-subdomain than DISK-stratified. Furthermore, under the MPP fitting, DISK-Stratified underestimates $\tau^2$, whereas the estimation remains precise when DISK-Subdomain is fitted with MPP. However, DISK-subdomain underestimates $\sigma^2$ more egregiously than DISK-stratified. Consequently, there is uncertainty regarding which data partitioning scheme to rely on while providing inference with DISK on model parameters. In contrast, the collaborative sketching approach aggregates inference over the data sketching mechanism, resulting in robust and accurate point estimation, as well as 95\% credible intervals that cover the true parameters.

The MSPE values presented in Table~\ref{tab:sim_pred_diagnostics} exhibit some variation between DISK-Subdomain and DISK-Stratified, although it is less pronounced compared to the parameter estimation. DISK-Subdomain performs inferior to DISK-Stratified across full-GP, MPP, and NNGP fittings on each subset. The coverage of both approaches is approximately equal, with DISK-Stratified showing slightly smaller interval score and energy score. This finding aligns with \cite{guhaniyogi2023distributed}, which demonstrates DISK-Stratified to be the best strategy among alternatives (e.g., DISK-Subdomain) for partitioning data subsets, as it includes representative samples from every subdomain in each subset, resulting in better model fitting using each subset posterior.
Notably, the collaborative sketching strategy yields a slightly smaller MSPE than DISK-Stratified. %Figure~\ref{fig:surface_plots} depicts the true response surface across $n_0$ out-of-sample locations for a randomly selected simulation. The estimated response surfaces for the collaborative sketching framework under full-GP, MPP, and NNGP illustrate the effective capture of the local features present in the true surface. 
While offering similar coverage, interval score, and energy score with DISK-Stratified for full-GP and NNGP, collaborative sketching provides better coverage and smaller interval and energy scores when sketched posteriors are fitted with MPP. In summary, the empirical investigation demonstrates the excellent performance of the collaborative sketching framework in terms of both inference and prediction when fitting each sketched posterior with a full-GP and its most popular computationally efficient alternatives.

%\begin{align}\label{eq: cov_fn}
%C(\bu,\bu';\theta_j) = \delta^2_j \exp\left\lbrace -\frac{1}{2} \left( \frac{||\bu-\bu'||}{\phi_j}\right) \right\rbrace,\:\:j=1,2,3,
%\end{align}
%with the true values of $\delta_1^2,\delta_2^2,\delta_3^2$ set to $1, 0.8, 1.1$, respectively. We fix the true values of $\phi_1,\phi_2,\phi_3$ at $1, 1.25, 2$, respectively.
%\begin{figure}[h!]
%\hspace{0.2cm}
%\begin{subfigure}{12cm}
%  \centering
%  \includegraphics[scale=0.35]{figures/true_surface.pdf}
%  \caption{True surface}
%\end{subfigure} \\ %
%\begin{subfigure}{.30\textwidth}
%  \centering
%  \includegraphics[width=.8\linewidth]{figures/sketching_full_gp.pdf}
%  \caption{Full GP, Sketching}
%\end{subfigure}
%\begin{subfigure}{.30\textwidth}
%  \centering
%  \includegraphics[width=.8\linewidth]{figures/sketching_mpp.pdf}
%  \caption{MPP, Sketching}
%\end{subfigure}
%\begin{subfigure}{.30\textwidth}
%  \centering
%  \includegraphics[width=.8\linewidth]{figures/sketching_sparse_gp.pdf}
%  \caption{NNGP, Sketching}
%\end{subfigure}
%\caption{The true surface plot of $y_s(\bu)$ at the 1,000 testing locations for a randomly selected test simulation, compared to the corresponding estimated surface plots under each model from the collaborative sketching approach.}
%    \label{fig:surface_plots}
%\end{figure}

\subsection{Choice of the Rank $m$ of Sketching Matrices}\label{sec:dim_m}
\begin{figure}
    \centering
    \includegraphics[scale=0.5]{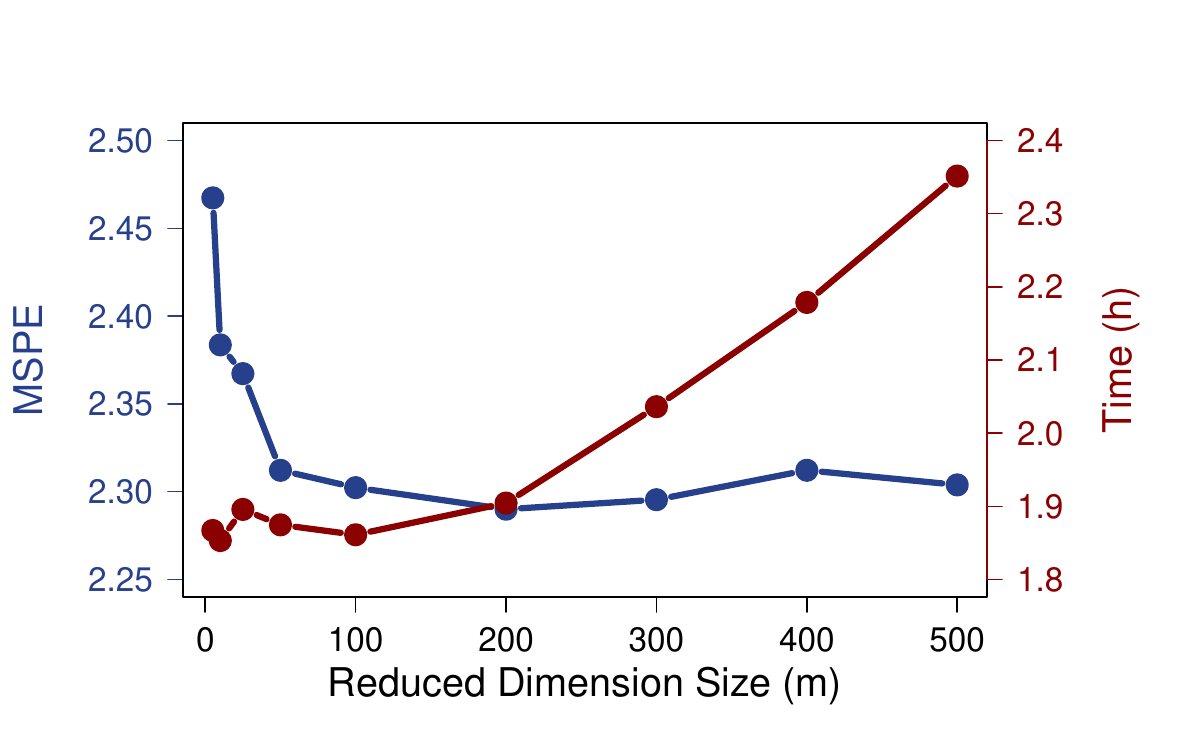}
    \caption{MSPE and computation time (in hours) for model fit and prediction in our simulation data, as a function of the compressed dimension size $m$.}
    \label{fig:mspe_by_dim}
\end{figure}

We explore the selection of an appropriate compression matrix size, denoted as $m$. The theoretical consideration suggests that $m$ should be on the order of $n/\log(n)$. However, in practical applications, it is feasible to achieve robust and accurate inference with smaller values of $m$. To illustrate this with simulated data having a sample size of $n=10,000$, we conducted model runs for various values of $m$. As expected, the MSPE decreases as $m$ increases, albeit with a diminishing rate of decline (refer to Figure~\ref{fig:mspe_by_dim}) until $m$ reaches 1\% of the full sample size $n$. Beyond this point, the MSPE stabilizes, a trend observed in our empirical experiments. Further increasing $m$ may result in marginal reductions in MSPE but comes at the expense of longer computation times (see Figure~\ref{fig:mspe_by_dim}). Therefore, for the real data analysis, and in particular, for SLOSH emulator data, we recommend utilizing $m$ as 1\% of the total sample size in the subsequent sections.

\section{SLOSH Emulator Data Analysis}\label{data_analysis}
We fit the collaborative sketching approach with full Gaussian process on the SLOSH emulator data described in Section~\ref{slosh}. As noted in Sections~\ref{sec:intro} and \ref{slosh}, this article offers the first principled functional analysis of SLOSH emulator data incorporating global and local attributes. Out of $10$ simulated storms each on $n=49719$ locations, $5$ storms are randomly selected for model fitting and rest are employed to assess predictive performance of the proposed approach. Following the discussion in Section \ref{sec:dim_m}, $m$ is set at 1\% of $n$, i.e., $m=498$. 

We compare our approach with Bayesian adaptive spline surfaces (BASS) \cite{francom2020bass}, a commonly used approach in national laboratories for analyzing functional data from multiple simulations. BASS is not able to incorporate local attributes (i.e., elevation in the SLOSH data) and is susceptible to high variance of estimation when the number of simulations is as low. However, given its popularity among researchers in national laboratories for the functional data analysis, we include it as a competitor and implement it using the \texttt{R} package \texttt{BASS}. Note that in this article, our primary focus centers on distributed Bayesian approaches. However, for benchmarking purposes, we also introduce a non-distributed method. Spatial statistics methods typically rely on a single simulation, making it challenging to compare our approach with an appropriate non-distributed state-of-the-art spatial techniques. In addressing this challenge, we employ the non-distributed NNGP on the complete dataset \cite{datta2016hierarchical} as a comparative method, albeit with a slight modification. Due to constraints in the \texttt{NNGP} package that allow only independent simulation runs, we perform predictive inference on NNGP by averaging results across multiple simulation runs. Despite the potential suboptimal performance of NNGP under this approach, we designate it as a benchmark to evaluate the non-distributed spatial model's performance across the entire dataset, referring to it as NNGP-ind.
%averaging over different simulation runs
%Similar to simulation studies, BASS is used as a competitor. Additionally, NNGP \cite{datta2016hierarchical} is employed as a competitor. Given that the \texttt{NNGP} package only allows the different simulations to be run independently, we draw predictive inference on NNGP 
%averaging over different simulation runs. We refer to the competitor as NNGP-ind.

\begin{figure}
    \centering
    \includegraphics[scale=0.65]{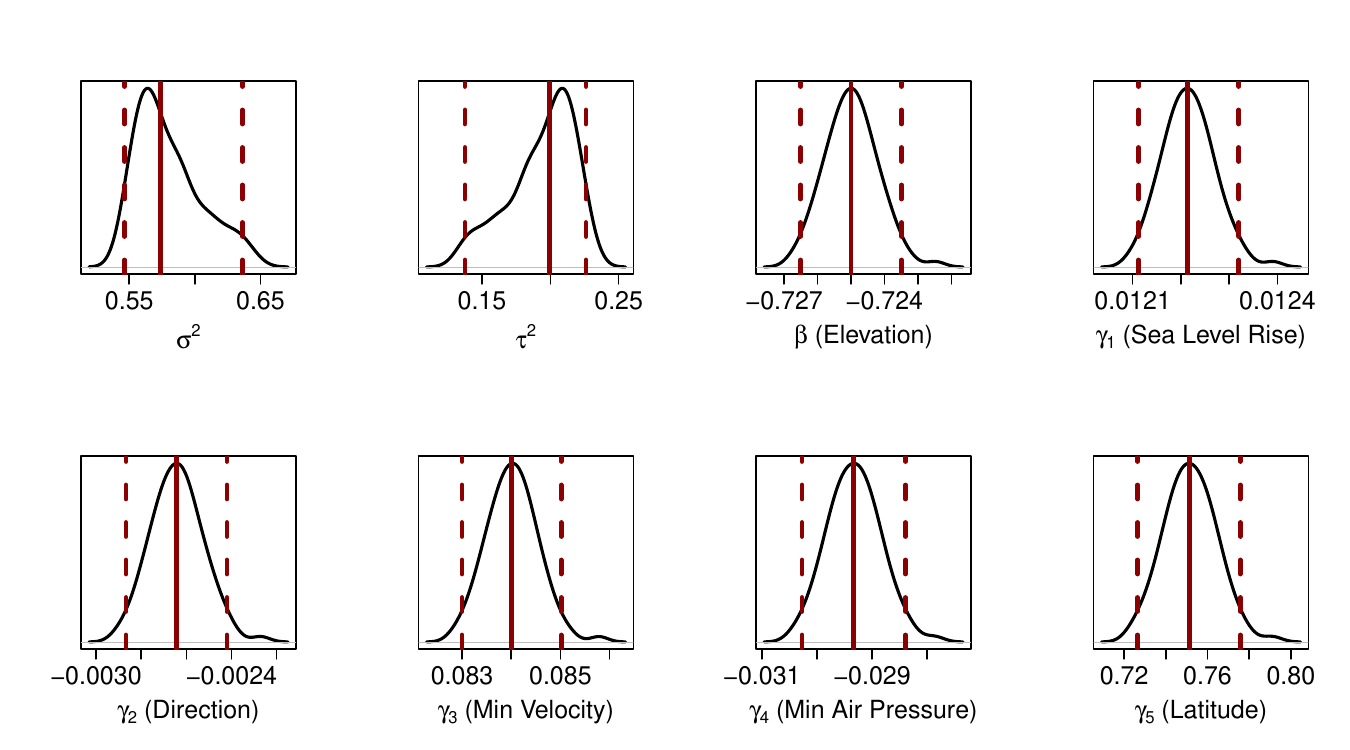}
    \caption{Densities for each parameter under our sketching model. Solid maroon lines indicate the posterior means, and dotted maroon lines indicate the upper and lower 95\% credible interval bounds.}
    \label{fig:densities}
\end{figure}

Figure~\ref{fig:densities} presents densities of the posterior distribution of model parameters with posterior median and 95\% CIs marked within each figure. All global and local attributes turn out to be significantly associated with the storm surge with none of their 95\% CIs includes zero. Understandably, the storm surge is positively associated with the sea level rise, minimum velocity of the eye of the storm, and negatively associated with the elevation, direction of the wind and minimum air pressure of the eye of the storm. The estimated posterior median of spatial variance $\sigma^2$ is dominant over the error variance $\tau^2$, justifying the spatial analysis.

\begin{figure}[h]
\begin{subfigure}{1.0\textwidth}
  \centering
  \hspace{0.5cm}
  \includegraphics[width=6cm]{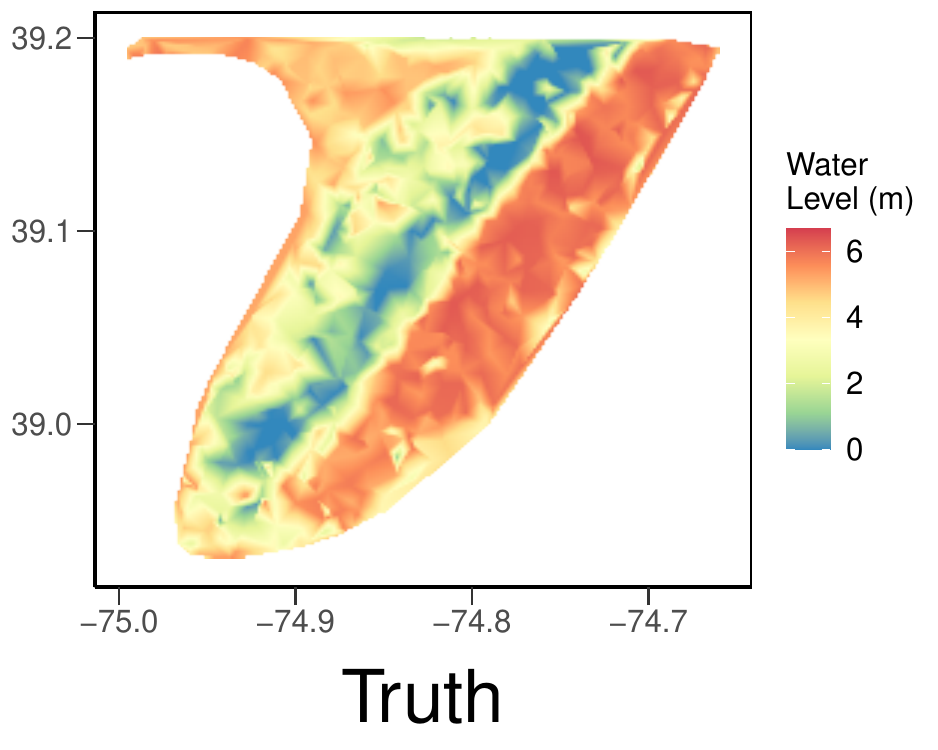}
\end{subfigure}  \\
\begin{subfigure}{1.0\textwidth}
  \centering
  \includegraphics[width=12cm]{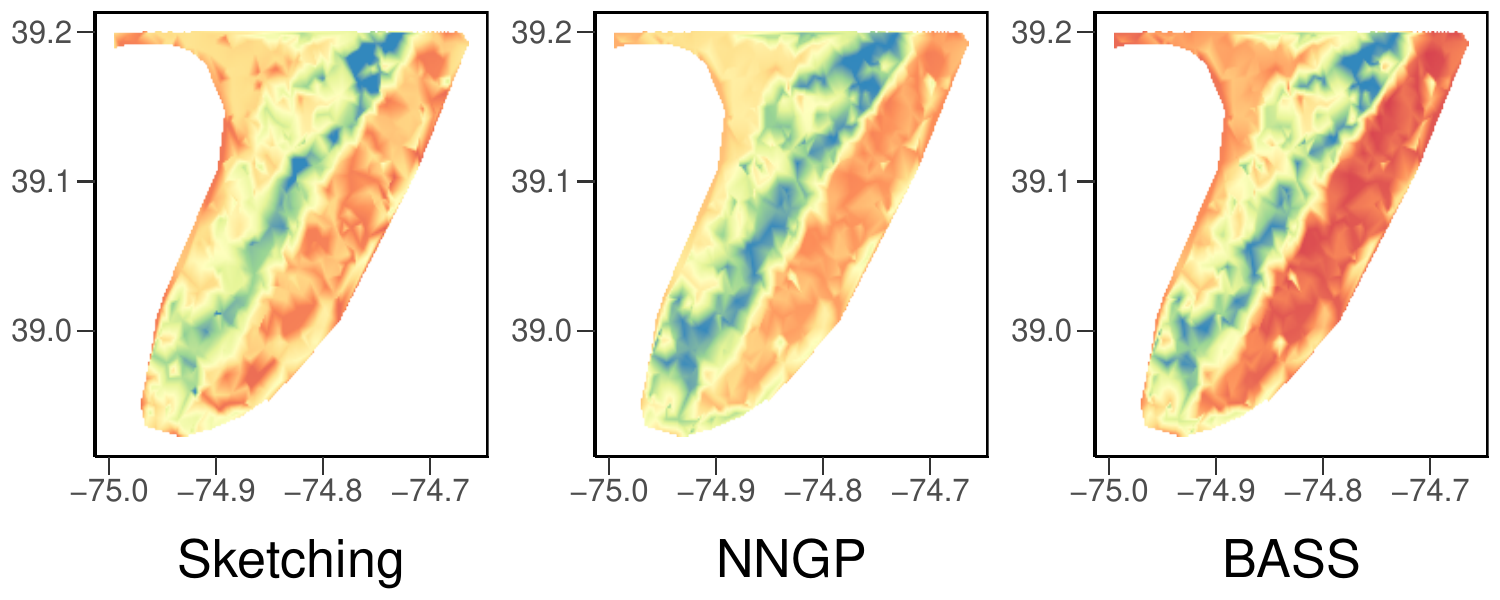}
\end{subfigure} \\ 
\caption{Actual water level (in meters) for a randomly selected storm at each coordinate in the testing dataset, along with the predicted water level under each model.}
\label{fig:storm10}
\end{figure}

\begin{table}[]
    \small
    \centering
    \begin{tabular}{rccccc}
    \toprule[1.5pt]
         & MSPE & Error \% & Coverage & Score & Energy score \\ 
         \midrule
         Sketching  & 1.07 & 0.06 & 0.88 & 5.60 & 0.63 \\
         NNGP-ind       & 0.83 & 0.09 & 0.18 & 20.45 & 1.52 \\
         BASS       & 1.45 & 0.10 & 0.59 & 11.23 & 1.02 \\
         %BART       & 0.72 & 0.08 & 0.83 & 5.97 & 0.48 \\
    \bottomrule[1.5pt]
    \end{tabular}
    \caption{Predictive diagnostics for the storm surge analysis. For interval score and Energy score, lower values indicate better scores.}
    \label{tab:storm_estimates}
\end{table}
Table~\ref{tab:storm_estimates} presents predictive inferences for all the competing approaches. The actual storm surge and the predicted storm surge in a randomly selected test storm in Figure~\ref{fig:storm10} demonstrate that the collaborative sketching approach, along with its competitors, accurately captures the local features of storm surge over space. Moreover, the results from Table~\ref{tab:storm_estimates} suggest that NNGP-ind might outperform other methods in terms of point predictions, as evidenced by its lowest MSPE. However, when considering the energy score, which utilizes predictive samples rather than the mean, the sketching approach is substantially favored over the NNGP-ind model. Therefore, while NNGP-ind provides excellent mean predictions, the sketching approach offers superior uncertainty quantification in prediction. This observation is further supported by the coverage and interval scores, where sketching exhibits significantly higher predictive coverage than its competitors, coupled with a much smaller interval score. Understandably, BASS underperforms both in terms of point estimation and uncertainty quantification due to the small number $S=5$ simulations for model fitting.

Section~\ref{slosh} highlights the significance of a flooding threshold set at four feet, considering its real-world implications, particularly in the context of power stations designed to withstand this level of floodwater. Given its importance, it is crucial for an emulator to accurately predict flood levels above four feet. To evaluate emulators in terms of this feature, we will examine the percentage of predictions that mis-classify the need for intervention, i.e., the percentage of predictions in the test simulations where the true SLOSH output exceeds four feet but the predicted output is less than four feet, or, the predicted output is above four feet and the true SLOSH output is less than four feet. Error percentage in Table~\ref{tab:storm_estimates} presents this metric for all the competitors. Notably, all competitors show small mis-classification rate with the rate being minimum for the collaborative sketching approach.

\section{Conclusion and Future Work}\label{conclude}
This article introduces a novel distributed Bayesian inferential framework that utilizes the theory of data sketching through random compression matrices. The article proposes fitting a powerful functional data model on multiple random sketches of the full data constructed using multiple random sketching matrices in parallel, followed by combining these inferences in a central server. By aggregating inference across diverse random sketches, our approach proves resilient to the selection of data sketches, leading to the development of novel robust distributed Bayesian learning approach. The proposed framework allows joint analysis of data stored within different research centers without leaking the privacy of samples. The proposed framework offers accurate inference on the association between water surge with storm-specific characteristics in the SLOSH simulator data. 

Our framework's ample generality enables its application in scaling complex data models. An immediate avenue for future work involves extending the framework to facilitate robust distributed inference with multivariate functional data models. Additionally, we plan to broaden the framework to support distributed Bayesian inference with Gaussian Cox process models, specifically tailored for large point process data.

\section{Acknowledgements}
Rajarshi Guhaniyogi acknowledges funding from National Science Foundation through DMS-2220840 and DMS-2210672; and funding from National Institute of Health Award Number R01NS131604.